\documentclass{article}
\usepackage{spconf,amsmath,graphicx}
\usepackage{amssymb}
\usepackage{multirow}
\usepackage{algorithmic}    % algorithm
\usepackage[linesnumbered,ruled]{algorithm2e} % algorithm
\usepackage{setspace}
\usepackage{cite}
\usepackage{enumitem} % iterm
\usepackage[colorlinks,
linkcolor=red,       %%修改此处为你想要的颜色
anchorcolor=red,  %%修改此处为你想要的颜色
citecolor=green,        %%修改此处为你想要的颜色，例如修改blue为red
]{hyperref}

\title{SRP-DNN: Learning Direct-Path Phase Difference for Multiple Moving Sound Source Localization}
\name{Bing Yang$^1$, Hong Liu$^1$, Xiaofei Li$^2$\thanks{This work is supported by National Natural Science Foundation of China (No.\,62073004),
Science and Technology Plan of Shenzhen (No.\,JCYJ20200109140410340).}}
\address{$^1$Key Laboratory of Machine Perception, Shenzhen Graduate School, Peking University, China \\
  $^2$Westlake University \& Westlake Institute for Advanced Study, Hangzhou, China\\
  {\normalsize \texttt{bingyang@sz.pku.edu.cn, hongliu@pku.edu.cn, lixiaofei@westlake.edu.cn}}
  }
\begin{document}
\ninept
\maketitle
\begin{abstract}
Multiple moving sound source localization in real-world scenarios remains a challenging issue due to interaction between sources, time-varying trajectories, distorted spatial cues, etc. In this work, we propose to use deep learning techniques to learn competing and time-varying direct-path phase differences for localizing multiple moving sound sources. A causal convolutional recurrent neural network is designed to extract the direct-path phase difference sequence from signals of each microphone pair. To avoid the assignment ambiguity and the problem of uncertain output-dimension encountered when simultaneously predicting multiple targets, the learning target is designed in a weighted sum format, which encodes source activity in the weight and direct-path phase differences in the summed value. The learned direct-path phase differences for all microphone pairs can be directly used to construct the spatial spectrum according to the formulation of steered response power (SRP). This deep neural network (DNN) based SRP method is referred to as SRP-DNN. The locations of sources are estimated by iteratively detecting and removing the dominant source from the spatial spectrum, in which way the interaction between sources is reduced. Experimental results on both simulated and real-world data show the superiority of the proposed method in the presence of noise and reverberation. % 100-150
\end{abstract}
\vspace{-0.1cm}

\vspace{-0.2cm}
\section{Introduction}
\label{sec:intro}
\vspace{-0.2cm}
Sound source localization aims to determine the relative position of sound sources with respect to microphone array.
As an important characteristics of directional sources, location information is widely used in real-world applications such as human-robot-interaction, and signal processing tasks including speech enhancement and source separation \cite{SESSOverview18}.
Recently, more and more works focus on localization in practical noisy and reverberant scenarios, %which faces the problem of interaction between sources, time-varying trajectories and distorted spatial cues.
but most of them either localize single moving source which avoids interaction between sources, or localize multiple static sources using long-time microphone signals. The dynamic trajectories of multiple moving sound sources pose new challenges to this task, which needs to timely estimate the locations of competing sources for each required time.
 %
%The location of moving sources can be block-wisely estimated using the methods designed for static sound sources or extend them for adaptive estimation.
%The position of source is treated as static in each single time block, and is estimated using a short sequence of signals.

Traditional methods, such as generalized cross correlation (GCC) \cite{GCC76}, steered response power (SRP) \cite{Yook16} and multiple signal classification (MUSIC) \cite{MUSIC86}, are widely used for sound source localization. To deal with the multi-source case, these methods are sometimes combined with time-frequency (TF) processing \cite{Griffin13-SCL,Wang16-SCL,Yan17,YBTASLP19}, where the W-disjoint orthogonal (WDO) assumption \cite{WDO04} is used to simplify the problem of multiple source localization on broadband to that of single source localization in individual TF bins.
Recently, deep learning based methods have shown promising localization performance \cite{CNN19, CCF17, SCL_CNN20,CRNN19, YBICASSP21, YBTASLP21}.  %due to the strong modelling ability of deep neural network (DNN).
They usually treat the localization task as either a location classification \cite{CNN19,CCF17,SCL_CNN20} or a location/feature regression \cite{CRNN19, YBICASSP21, YBTASLP21} problem.
%These methods can be classified into three categories according to the DNN output, namely location classification, location regression and feature regression \cite{}.
Location classification methods output the posterior probabilities of location classes, but the dimension of output will increase with the number of candidate locations.  %especially for joint azimuth and elevation estimation.
Regressing the locations or features of multiple sources is difficult, because the assignment between multiple outputs and multiple training targets become ambiguous, and the dimension of regression output ought to vary with the number of sound sources. %Hence, a learning method is required.

%Among them, regressing localization cues from distorted signals or features preserves the locally monotonic correlations between features and locations.
%It provides a simple and effective way to improve the robustness of localization under adverse conditions. %and generalize to unseed dual-microphones arrays.

%The block-wise methods consider the spatial information of short-term context, and process each time block independently. Different from these methods,
Based on the fact that sound sources move continuously over time, many works try to exploit temporal context for moving sound source localization \cite{LXF19, CRNN19,Cross3D21,TempContext21}.
%based on the signal correlations of consecutive frames.
%To deal with the moving source case, the temporal estimates are smoothed by particle filter, kalman filter, etc.
Diaz-Guerra \emph{et al}. performed a three-dimensional (3D) convolutional neural network (CNN) over the sequence of SRP-phase transform (SRP-PHAT) spatial spectrum %dimensions of SRP with phase transform (SRP-PHAT) power maps and the temporal dimension
to track single source \cite{Cross3D21}.
Li \emph{et al}. recursively estimated the direct-path relative transfer function (DP-RTF) using a short memory for online multiple-speaker localization \cite{LXF19}.
%Bohlender \emph{et al}. incorporated the long-term temporal context into the CNN based multi-source DOA estimation \cite{TempContext21}.
Despite the progress of these works, %localization of moving sources still remains a challenging problem.
it still requires a method to %remains to be investigated that
%how to
well exploit temporal context to quickly detect multiple moving sources and meanwhile filter out the outlier estimates caused by noise and reverberation.

This paper works on taking full use of the spatial-temporal context information to localize multiple moving sound sources. Considering the localization robustness %and the temporally continuous characteristic
of direct-path features \cite{YBICASSP21,YBTASLP21}, the sequence of direct-path inter-channel phase differences (IPDs) is predicted by deep neural network (DNN) for each microphone pair.
These predicted IPDs of all microphone pairs are used to construct the SRP spatial spectrum. This improved SRP is referred to as SRP-DNN.
%The direction of arrivals (DOAs) of active sources are estimated from the improved spatial spectrum for each required time.
%To online estimate, the SRP-PHAT is constructed, from which DOA of sources are estimated.
%The whole block diagram is shown in Fig.~\ref{fig:meth_flowchart} (a).
%A moving sound source localization method with temporal direct-path
%feature is designed, and its block diagram is shown in Fig.~\ref{fig:meth_flowchart}. The temporal direct-path phase feature is first predicted by causal CRNN, and then used to construct the SRP-PHAT spatial spectrum and estimate DOA of sources. %The DOA of sources is determined by iteratively detecting and localizing sound sources.
%and use them to localize sources.
%Note that we do not deal with the data association problem in this work.
%This work has the following contributions.
The contributions of this work are summarised as follows.

\textit{Learning direct-path IPD sequence for multiple moving sources:}
A causal convolutional recurrent neural network (CRNN) is designed to predict the direct-path IPD sequence for each microphone pair. This architecture fully exploits the TF patterns of direct-path IPDs, such as the temporal smoothness due to the continuous movement of sound sources, and the linearity along frequencies due to the linear relation between IPD and time difference of arrival (TDOA).
The magnitude and phase spectrograms of dual-channel microphone signals are taken as the network input.
As for the network output, it is difficult to well separate the direct-path IPDs of different sources from the overlapped microphone signals.
%An online version can be implemented due to the causality of this model.
%To avoid the inconvenience of simultaneously predicting the features of multiple sources,
Hence, the learning target is designed in a weighted sum format, where the weight reflects the source activity and the summed value compounds the direct-path IPDs. Importantly, the learned weighted direct-path IPDs can be directly used to construct the spatial spectrum for multiple sources.
%The proposed method avoids parameter increasing of location classification with increasing candidate spatial directions. %to determine the azimuth and elevation of sources
%As the feature dimension increases with the number of microphones and so does the number of parameters, we use mic-pair-sharing feature estimation network.

\begin{figure*}[t]
  \centering
  \includegraphics[width=0.9\linewidth]{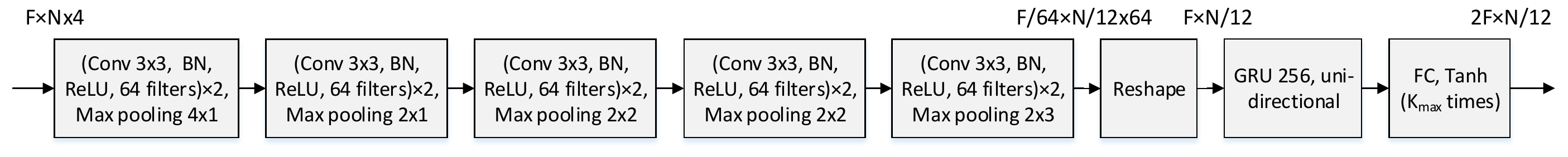}
  \vspace{-0.35cm}
  \caption{%Flowchart of the proposed moving sound source localization method.
  Causal CRNN architecture to estimate the sequence of direct-path phase difference. }
  \label{fig:meth_net}
  \vspace{-0.3cm}
\end{figure*}

\textit{Iterative source detection and localization:}
One trivial way to perform direction of arrival (DOA) estimation is directly applying the peak detection method \cite{Yan17} on the spatial spectrum , which however is inaccurate especially for the case where multiple peaks are merged due to the interaction between sources. %when multiple sources are close to each other.
To solve this problem, this work proposes a new method that iteratively detects and removes the dominant source from the overall spatial spectrum, which makes it possible to separate the merged peaks.

\vspace{-0.2cm}
\section{Method}
\vspace{-0.2cm}
Suppose that there are multiple moving sound sources in the noisy and reverberant environment, the signal captured by the $m$-th microphone is formulated in the short-time Fourier transform domain as:
%\begin{equation}
%    x_m(t) = \sum_{k=1}^{K} h_{m}(t, \boldsymbol{\theta}_k) * s_k(t) + v_m(t),
%  \label{eq_model_t}
%\end{equation}
\vspace{-0.2cm}
\begin{equation}
    X_m(n,f) =  \sum_{k=1}^{K} H_m(n, f, \boldsymbol{\theta}_k) S_k(n,f)  + V_m(n,f),
    \label{eq_model_tf}
    \vspace{-0.16cm}
\end{equation}
where $m$$\in$$[1,M]$, $k$$\in$$[1,K]$ $n$$\in$$[1, N]$ and $f$$\in$$[1, F]$ are the indices of  microphones, sound sources, time frames and frequencies, respectively.
$\boldsymbol{\theta}_k$$=$$[\theta_k^{\rm ele}, \theta_k^{\rm azi}]^T$
%($\theta_k^{\rm ele}\in [0,\pi]$,$\theta_k^{\rm azi}\in [-\pi,\pi)$)
denotes the 2D DOA
%(both elevation and azimuth)
of the $k$-th sound source, which consists of elevation $\theta_k^{\rm ele}$$\in$$[0,\pi]$ and azimuth $\theta_k^{\rm azi}$$\in$$[-\pi,\pi)$. Here, $(\cdot)^T$ denotes matrix/vector transpose.
$\theta_k^{\rm ele}$$=$$0$ and $\theta_k^{\rm azi}$$=$$0$ are defined along the positive $z$-axis and the positive $x$-axis, respectively.
$X_m(n,f)$, $S_k(n,f)$ and $V_m(n,f)$ represent the microphone, source and noise signals, % in the TF domain,
respectively. $H_m(n, f, \boldsymbol{\theta}_k)$ denotes the time-varying acoustic transfer function from the $k$-th source to the $m$-th microphone.

\vspace{-0.2cm}
\subsection{Classical SRP-PHAT }
\vspace{-0.1cm}

The classical SRP-PHAT algorithm \cite{LOCATA_SRP} estimates the spatial spectrum  by averaging the frame-wise GCC-PHAT over frequencies and nonredundant microphone pairs, namely:
\vspace{-0.2cm}
\begin{equation}
    P(\boldsymbol{\theta},n) = \frac{2}{M(M-1)}  \sum_{m=1}^{M-1}\sum_{m'=m+1}^{M}  %\sum_{n \in \mathbb{S}_l}
    {G_{mm'}(\boldsymbol{\theta},n)},
    \label{eq_SRP-PHAT}
    \vspace{-0.16cm}
\end{equation}
%where $l\in [1,L]$ denotes the index of time interval, and $\mathbb{S}_l$ represents the set of time frame indexes belongs to the $l$-th time interval.
where $\boldsymbol{\theta}$ denotes the candidate DOA for spatial spectrum construction. %$G_{mm'}(\boldsymbol{\theta},n)$ is
The frame-wise GCC-PHAT \cite{Mask_RNN19,GCC21} for one pair of microphone signals is computed as:
\vspace{-0.2cm}
\begin{equation}
    G_{mm'}(\boldsymbol{\theta},n) = \frac{1}{F} \sum_{f=1}^F  \mathfrak{R}\left\{ \Psi(n,f)
    e^{-j\omega_f \tau_{mm'}(\boldsymbol{\theta})} \right\},
   % =&[\cos\angle{X_m(n,f)X_{m'}^{*}(n,f)},\sin\angle{X_m(n,f)X_{m'}^{*}(n,f)}]\\ &[\cos\omega_f\tau_{mm'}(\boldsymbol{\theta}),\sin\omega_f\tau_{mm'}(\boldsymbol{\theta})]^T,
    \label{eq_GCC-PHAT}
    \vspace{-0.16cm}
\end{equation}
with the PHAT cross-power spectrum being:
\vspace{-0.13cm}
\begin{equation}
    \Psi(n,f) = \frac{X_m(n,f)X_{m'}^{*}(n,f)}{\left|X_m(n,f)X_{m'}^{*}(n,f)\right|}=e^{j\angle{X_m(n,f)X_{m'}^{*}(n,f)}},
    \label{eq_PHAT}
    \vspace{-0.13cm}
\end{equation}
where $\omega_f$ denotes the angular frequency of the $f$-th frequency.
%$\tau_{mm'}(\boldsymbol{\theta})$ is the TDOA between signals captured by the $m$-th and the $m'$-th microphones, which can be calculated based on the far-field acoustic propagation model.
$\mathfrak{R}\{\cdot\}$ and $|\cdot|$ denote the real part and magnitude of complex number, respectively.
For far-field model where the propagation paths from one sound source to different microphones are regarded to be parallel, the TDOA between signals captured by the $m$-th and the $m'$-th microphones is computed as:
\vspace{-0.1cm}
\begin{equation}
    \tau_{mm'}(\boldsymbol{\theta}) =  {d_{mm'}(\boldsymbol{\theta})}/{c} = {(\mathbf{p}_{m}-\mathbf{p}_{m'})^T \mathbf{u}(\boldsymbol{\theta})}/{c},
     %\tau_{m'}(\boldsymbol{\theta})- \tau_{m}(\boldsymbol{\theta}).
  \label{eq_tdoa}
  \vspace{-0.1cm}
\end{equation}
with the Cartesian coordinates of the direction $\boldsymbol{\theta}$ on a unit sphere being:
\vspace{-0.1cm}
\begin{equation}
    \mathbf{u}(\boldsymbol{\theta}) =
    [\sin(\theta^{\rm ele})\cos(\theta^{\rm azi}),
    \sin(\theta^{\rm ele})\sin(\theta^{\rm azi}),
    \cos(\theta^{\rm ele})]^T,
  \label{eq_u}
  \vspace{-0.1cm}
\end{equation}
where $\mathbf{p}_{m}$ denotes the location coordinate of the $m$-th microphone and $c$ is the speed of sound.
%Clearly, the classical SRP-PHAT discards the magnitude information of the cross-power spectrum, and estimates the spatial spectrum of each microphone pair only based on the phase of the cross-power spectrum, namely the IPD.

\vspace{-0.2cm}
\subsection{Direct-Path Phase Difference Learning}
\vspace{-0.1cm}
For the noise-free and anechoic single-source case,
%the IPD is only related to the direct-path signals, and
the PHAT cross-power spectrum, i.e. Eq.~\eqref{eq_PHAT}, is actually computing the complex-valued direct-path IPD for the source at $\boldsymbol{\theta}_k$:
\vspace{-0.1cm}
\begin{equation}
    \Psi^{\boldsymbol{\theta}_k}(n,f) =e^{j\omega_f \tau_{m m'}(\boldsymbol{\theta}_k)}.
    \label{eq_dp_PHAT}
    \vspace{-0.1cm}
\end{equation}
Accordingly, the value of the frame-wise GCC-PHAT ranges from 0 to 1, and reaches the maximum when $\boldsymbol{\theta}$=$\boldsymbol{\theta}_k$, which is also the case for Eq.~\eqref{eq_SRP-PHAT}. %, which is the same for the SRP-PHAT based spatial spectrum (Eq.~\eqref{eq_SRP-PHAT}).
%The spatial spectrum based on SRP-PHAT tends to show a sharp peak at the actual DOA. %only depends on TDOAs of microphone pairs, which
%DOA can be estimated by detecting the highest peak of spatial spectrum, namely:
%\vspace{-0.1cm}
%\begin{equation}
%    \hat{\boldsymbol{\theta}}_k = \arg\max_{\boldsymbol{\theta}} P(\boldsymbol{\theta},n).
%    \label{eq_SSL}
%    \vspace{-0.1cm}
%\end{equation}
However, ambient noise and room reverberation are inevitable in realistic applications.
%Using Eq.~\eqref{eq_PHAT}, the spatial spectrum peaks associated with the direct-path signals may be altered, overlapped or attenuated by noise, reflected paths or interfering sources due to the contaminated direct-path IPDs, which will lead to DOA estimation errors.
In these cases, using Eq.~\eqref{eq_PHAT},  the direct-path IPD is contaminated by noise, reverberation and interfering sources, which will no doubt lead to less prominent peaks of spatial spectrum and thus DOA estimation error.
Therefore, recovering the direct-path IPDs from the noisy and reverberant microphone signals is essential for the SRP-PHAT based localization methods.
%As direct-path features is an important spatial cue of sources, and noise and reverberation tends to introduce distortion,%
%In the anechoic and noiseless case, sensor signals only involve direct-path propagation.
%the ideal value for the phase ought to only dependent on the direct-path signals.

In this study, we propose to leverage the strong modeling ability of DNN to learn the direct-path IPDs for each microphone pair. Considering the continuous moving properties of sources over time, the TF patterns of direct-path phase differences are exploited and modelled by a causal CRNN. %for the direct-path feature learning.
The convolutional units capture the inter-channel information and its short-term temporal context, %to track the changed spatial information with the movement of sources,
and the recurrent units exploit the long-term spatial-temporal context. %to filter out the contamination of acoustic interferences.
Overall, the CRNN aims to filter out the contamination of acoustic interferences and recover the direct-path IPDs for multiple moving sources.
The causality of this model facilitates the online implementation of sound source localization.
The details are described as follows.

\textit{Network architecture:}
%The architecture of causal CRNN is shown in Fig.~\ref{fig:meth_network}
Signals recorded by different microphone pairs are treated as independent training samples, and they are processed separately by the proposed network during inference time.
The logarithm-magnitude and phase spectrograms of dual-channel signals are taken as the input features of the CRNN. As shown in Fig.~\ref{fig:meth_net}, the input features are passed to ten causal convolutional modules. Each module consists of a causal convolutional layer followed by a batch normalization (BN) and a rectified linear unit (ReLU) activation function. %These causal convolutional layers are with 64  3 $\times$ 3 kernels.
A max pooling is used to compress the frequency and time dimensions after each two convolutional modules.
The output of CNN layers is flatten for the frequency and filter dimensions, and then fed into one-layer uni-directional gated recurrent unit (GRU) with 256 hidden units.
A  fully connected (FC) layer with an activation of $K_{\rm max}$ times tanh function is used to output the direct-path phase difference (See details in \emph{Learning target})
%summed direct-path IPD vector (defined in Eq.~\eqref{eq_target})
for one microphone pair.
Here, $K_{\rm max}$ refers to the maximum possible number of sources. %The number of time frames is compressed from $N$ to $N/12$ after passing to the causal CRNN.
Note that, the frame rate of microphone signals (about 60 frames per second) is normally too high relative to the need of localization frame rate (about 5 frames per second is enough). Therefore, the input frame rate is compressed by a factor of 12 at the network output, and $n'$ is used to denote the frame index of output.
%One output frame corresponds 12 input frames.

\textit{Learning target:}
%\subsection{Causal CRNN}
To learn the direct-path IPD (or its complex-valued form) for the single-source case, a simple way is to directly regress the real and imaginary parts of Eq. \eqref{eq_dp_PHAT}, which is expressed in vector form with all frequencies as:
\vspace{-0.15cm}
\begin{equation}
\begin{aligned}
   \mathbf{r}_{mm'}(\boldsymbol{\theta}_k) &=  [\cos\left(\omega_1\tau_{mm'}(\boldsymbol{\theta}_k)\right),\sin\left(\omega_1\tau_{mm'}(\boldsymbol{\theta}_k)\right),\ldots,\\
   &\cos\left(\omega_F\tau_{mm'}(\boldsymbol{\theta}_k)\right),\sin\left(\omega_F\tau_{mm'}(\boldsymbol{\theta}_k)\right)]^T \in \mathbb{R}^{2F\times1}.
    \label{eq_target_single}
     \vspace{-0.15cm}
\end{aligned}
\end{equation}
%and $2F$ values needs to be estimated for each microphone pair, namely the sine and cosine of the full-band IPD.  %and hence the training target is $\mathbf{R}(\boldsymbol{\theta}_k)$.
For the case of multiple sources, %multiple representations needs to be estimated.
directly regressing multiple vectors will
require the dimension of regression output to be variant along with the variance of the number of sound sources. In addition, there will be the assignment ambiguity between the multiple outputs and the multiple training targets, which is similarly encountered by the speech separation task \cite{PIT17,PIT17T}.
%face the dependence of the number of model output on the number of sources (or at least the maximum number), and also the label ambiguity during model training.
To avoid these problems, we propose to add up the direct-path IPD vectors of multiple sound sources as the training target, which is formulated as:
%The vectors of two sources with enough angular distance can be separated by projection to the direct-path vectors of candidate DOAs, since they can be treated as uncorrelated. Considering the intermittently sounding properties of sources, the source-activity is encoded as the weight of direct-path vectors, and accordingly the learning target is:
\vspace{-0.22cm}
\begin{equation}
   \mathbf{R}_{mm'}(n') =  \sum_{k=1}^{K}{\beta_k(n') \mathbf{r}_{mm'}(\boldsymbol{\theta}_k)}.
    \label{eq_target}
     \vspace{-0.15cm}
\end{equation}
The summation weight $\beta_k(n')$ is defined as
the activity probability of the $k$-th source at the $n'$-th output frame, which is computed as, over the 12 input frames that corresponding to the $n'$-th output frame, the proportion of the input frames where the $k$-th source is active. It ranges from 0 to 1.
Correspondingly, the elements of the summed vector are in the range of $[0, K]$.
%the ratio of the sounding time in the current time, which is calculated based on the WebRTC voice activity detector \cite{vad19}.
%and $F$ frequency bins, totally $M(M-1)F$ needs to be estimated for each time.
The mean squared error (MSE) between the network output and the learning target is taken as the training loss.

This learning target is reasonable in the sense that taking the inner product between this summed vector $\mathbf{R}_{mm'}(n')$ (or its prediction) and the direct-path IPD vectors of candidate DOAs (denoted as $\mathbf{r}_{mm'}(\boldsymbol{\theta})$) is equivalent to the weighted summation of the direct-path spatial spectra (for one microphone pair) of multiple sources.  %directly constructed by
%, due to the local linearity of Eq.~\eqref{eq_GCC-PHAT}.
This can be interpreted by substituting Eq.~\eqref{eq_target} into the inner product,
namely:
%the GCC-PHAT spatial spectrum of multiple sources can be directly constructed by taking the inner product between this summed vector $\mathbf{R}_{mm'}(n')$ (or its prediction) with the IPD vector of a candidate DOA (denoted as $\mathbf{r}_{mm'}(\boldsymbol{\theta})$), namely: %due to the linearity of Eq.~\eqref{eq_GCC-PHAT},  as
\vspace{-0.22cm}
\begin{equation}
\begin{aligned}
    \mathbf{R}^{T}_{mm'}(n')\mathbf{r}_{mm'}(\boldsymbol{\theta})
    =&\sum_{k=1}^K \beta_k(n') \mathbf{r}^{T}_{mm'}(\boldsymbol{\theta}_k) \mathbf{r}_{mm'}(\boldsymbol{\theta})\\
    %=&\sum_{k=1}^K \beta_k(n')  \sum_{f=1}^F \mathfrak{R}\left\{\Psi^{\boldsymbol{\theta}_k}(n',f)e^{-j\omega_f \tau_{mm'}(\boldsymbol{\theta})} \right\} \\
    = &\sum_{k=1}^K \beta_k(n')FG^{\boldsymbol{\theta}_k}_{mm'}(\boldsymbol{\theta},n'),
    \label{eq_inner}
\end{aligned}
    \vspace{-0.15cm}
\end{equation}
with the direct-path spatial spectrum for the $k$-th source being:
\vspace{-0.2cm}
\begin{equation}
    G^{\boldsymbol{\theta}_k}_{mm'}(\boldsymbol{\theta},n') = \frac{1}{F}\sum_{f=1}^F \mathfrak{R}\left\{\Psi^{\boldsymbol{\theta}_k}(n',f)e^{-j\omega_f \tau_{mm'}(\boldsymbol{\theta})} \right\}.
    \label{eq_GCC-PHAT-DP}
    \vspace{-0.2cm}
\end{equation}
%Here, $\mathbf{r}_{mm'}(\boldsymbol{\theta})$ is the direct-path IPD vector of a candidate DOA, which acts as a template in the inner product based feature matching. $\mathfrak{R}\left\{\Psi^{\text{dp}}(n',f)e^{-j\omega_f \tau_{mm'}(\boldsymbol{\theta})} \right\}$ is exactly Eq.~\eqref{eq_GCC-PHAT} using direct-path signals.
%The Deep-SRP spatial spectrum $P^{\text{dp}}(\boldsymbol{\theta},n)$ is obtained by averaging the inner products in Eq.~\eqref{eq_GCC-PHAT-DPM} over different microphone pairs.
 %loss %between the network output and the learning target is used to train the designed model.
%\begin{figure}[t]
%  \centering
%  \includegraphics[width=0.5\linewidth]{../Figure/meth_network.pdf}
%  \vspace{-0.2cm}
%  \caption{Causal CRNN architecture for direct-path feature estimation.}
%  \label{fig:meth_network}
%  \vspace{-0.2cm}
%\end{figure}

\vspace{-0.35cm}
\subsection{DOA Estimation of Multiple Moving Sources}
\vspace{-0.1cm}
Using the prediction of the summed direct-path IPD vector, denoted as $\hat{\mathbf{R}}_{mm'}(n')$, the overall spatial spectrum of SRP-DNN is estimated with all microphone pairs as:
\vspace{-0.2cm}
\begin{equation}
\begin{aligned}
   &P'(\boldsymbol{\theta},n')=\frac{2}{M(M-1)F}  \sum_{m=1}^{M-1}\sum_{m'=m+1}^{M} \hat{\mathbf{R}}^{T}_{mm'}(n')\mathbf{r}_{mm'}(\boldsymbol{\theta}).
   % =& \sum_{k=1}^K \beta_k(n')\frac{2}{M(M-1)} \sum_{m=1}^{M-1}\sum_{m'=m+1}^{M} G^{\boldsymbol{\theta}_k}_{mm'}(\boldsymbol{\theta},n',f),  %\mathfrak{R}\left\{\Psi_{\boldsymbol{\theta}_k}^{\text{dp}}(n',f)e^{-j\omega_f \tau_{mm'}(\boldsymbol{\theta})} \right\},
    \label{eq_Deep_SRP}
     \vspace{-0.13cm}
\end{aligned}
\end{equation}
It is an estimation of the following theoretical value:
\vspace{-0.2cm}
\begin{equation}\nonumber
   \sum_{k=1}^K \beta_k(n')\frac{2}{M(M-1)} \sum_{m=1}^{M-1}\sum_{m'=m+1}^{M} G^{\boldsymbol{\theta}_k}_{mm'}(\boldsymbol{\theta},n'). %\mathfrak{R}\left\{\Psi_{\boldsymbol{\theta}_k}^{\text{dp}}(n',f)e^{-j\omega_f \tau_{mm'}(\boldsymbol{\theta})} \right\},
    \label{eq_Deep_SRP_}
     \vspace{-0.13cm}
\end{equation}
Since $\sum_{m=1}^{M-1}\sum_{m'=m+1}^{M} G^{\boldsymbol{\theta}_k}_{mm'}(\boldsymbol{\theta},n')$ exhibits a high peak at $\boldsymbol{\theta}_k$ and has a very small value at other directions, the value of $P'(\boldsymbol{\theta}_k,n')$ is  dominantly contributed by the source at $\boldsymbol{\theta}_k$, and is thus approximately equal to $\beta_k(n')$. %expected to have a approximated value $\beta_k(n')$ which is dominantly contributed by the $k$-th source.
%We utilize totally $M(M-1)/2$ nonredundant microphone pairs, and
%Using the prediction of the summed direct-path IPD vector, denoted as $\hat{\mathbf{R}}_{mm'}(n')$,

Multiple sources can be localized by the peak detection method \cite{Yan17} which searches the significant peaks of the estimated spatial spectrum that are larger than a predefined threshold,
%detecting the peaks of the spatial spectrum of,
as is done for regular SRP-PHAT.
%The direct-path features for all the $M(M-1)/2$ nonredundant microphone pairs are extracted using the same casual CRNN network.
%After that, the improved spatial spectrum is estimated by replacing the real and imaginary parts of Eq. \eqref{eq_PHAT} with that of these predicted direct-path features.
However, the peaks of sources may be merged due to the interaction between sources. To solve this problem, this work proposes to estimate the DOAs of active sources by iteratively detecting and removing the dominant source from the overall spatial spectrum, which is summarized in Algorithms 1. Since the iterative method works independently for each time step, the frame index $n'$  is omitted for simplicity.
The DOA of the dominant source $\hat{\boldsymbol{\theta}}^{\rm d}$ can be easily estimated as the candidate direction having the largest value of $P'(\boldsymbol{\theta})$. According to  Eq.~\eqref{eq_target}, the contribution of the dominant source can be removed by subtracting $\hat{\beta}^{\rm{d}} \times \mathbf{r}_{mm'}(\hat{\boldsymbol{\theta}}^{\rm d})$ from $\hat{\mathbf{R}}_{mm'}$, where $\hat{\beta}^{\rm{d}}$ is approximated by $P'(\hat{\boldsymbol{\theta}}^{\rm{d}})$.
%one need to determine its summation weight. The weight can be approximately assumed to be equal to  $P'(\hat{\boldsymbol{\theta}}^{\rm d},n')$, since this value is mainly contributed by the dominant source.
The new dominant source can be detected after the contribution of the previous one is removed. The iteration will stop when there is no notable source remained, namely $\hat{\beta}^{\rm{d}}<{\beta}_{\rm TH}$. Here, ${\beta}_{\rm TH}$ is a predefined threshold.

\begin{algorithm} [!t]
  %\algsetup{linenosize=\tiny}
  %\setstretch{0.1}
  %\small
  \footnotesize
  %\scriptsize
  \caption{Iterative source detection and localization.}
  \KwIn{The predicted direct-path IPD features $\{\hat{\mathbf{R}}_{mm'}\}$.} %the feature template for all candidate directions $\{\mathbf{r}(\boldsymbol{\theta})\}$.}
  \KwOut{ The DOA estimates $\{\hat{\boldsymbol{\theta}}_1, \ldots,\hat{\boldsymbol{\theta}}_k\}$.}
        %$i \leftarrow 0$; \qquad \quad \ $\rightarrow$ Initialization of iteration index \\
        \For {$k \gets 1$ to $K_{\rm max}$}
        {
            Estimate the spatial spectrum $P'(\boldsymbol{\theta})$ with $\{\hat{\mathbf{R}}_{mm'}\}$;\\
            Estimate the DOA of the dominant source:  $\hat{\boldsymbol{\theta}}^{\rm{d}}$$\gets$$\arg\max_{\boldsymbol{\theta}} P'(\boldsymbol{\theta})$;\\ %as the candidate direction that has the maximum value of $P(\boldsymbol{\theta})$;\\ %according to Eq. \eqref{eq_SSL}; \\
            Remove the contribution of the dominant source:\!
            $\hat{\beta}^{\rm{d}}$$\gets$$P'(\hat{\boldsymbol{\theta}}^{\rm{d}})$, \!and for all microphone pairs $\hat{\mathbf{R}}_{mm'}$$\gets$$\hat{\mathbf{R}}_{mm'}$$-$$\hat{\beta}^{\rm{d}}$$\times$$ \mathbf{r}_{mm'}(\hat{\boldsymbol{\theta}}^{\rm{d}})$;\\
            \If {$\hat{\beta}^{\rm{d}}<{\beta}_{\rm TH}$}
            {
                The dominant source at $\hat{\boldsymbol{\theta}}^{\rm{d}}$ is inactive, $k \gets k-1$,
                \textbf{break};
            }
            $\hat{\boldsymbol{\theta}}_k \gets \hat{\boldsymbol{\theta}}^{\rm{d}}$;
        }
        \Return $\{\hat{\boldsymbol{\theta}}_1, \ldots,\hat{\boldsymbol{\theta}}_k\}$.
\label{alg:IDL}
\end{algorithm}

\vspace{-0.2cm}
\section{Experiments and Discussions}
\vspace{-0.2cm}
\subsection{Experimental Setup}
\vspace{-0.1cm}

%\subsubsection{Dataset}
Multi-conditional microphone signals are generated for network training. Following the data generation procedure presented in \cite{Cross3D21}, the size of simulated rooms are randomly selected in the range from 3$\times$3$\times$2.5 m to 10$\times$8$\times$6 m, and the reverberation time is randomly set in the range from 0.2 s to 1.3 s. A 12-microphone array
is randomly placed inside the room, and the array geometry is set to be the same as that mounted on the NAO robot head in the localization and tracking (LOCATA) challenge dataset\cite{LOCATA18}.
The sound source moves along a random sinusoidal continuous trajectory.
According to these room settings, room impulse responses (RIRs) are generated using the image method \cite{RIR79} implemented by the gpuRIR \cite{gpuRIR20}.
Speech recordings are randomly selected from the LibriSpeech corpus \cite{LibriSpeech15}.
The microphone signals are created by filtering speech recordings by the RIRs, and then adding Gaussian noise with a SNR from 5 dB to 30 dB.
In order to increase the diversity of training
%temporal context and promote the adaptability of training model to varying
acoustic conditions,  %we increase the diversity of temporal context by a random data generation scheme.
%According to the physical principle of acoustic generation,
each training sample is generated on-the-fly as a random combination of data settings regarding source trajectories, microphone positions, source signals, noise signals, reverberation times, SNRs, etc.
%Hence, a infinite-size dataset is available for model training.
The evaluation is performed on both the simulated dataset and the LOCATA dataset. The real-world data provided by the LOCATA dataset is recorded in a computing laboratory with a size of 7.1$\times$9.8$\times$3 m. The reverberation time is 0.55 s.
We consider the development and evaluation sets of tasks 3 and 5 with a single moving source, and also that of tasks 4 and 6 with two moving sources for performance evaluation. %, each task comprising three recorded sequences.

%\subsubsection{Parameter setting and evaluation metric}
The sampling rate of microphone signals is 16 kHz. The window length and the frame shift are 32 ms and 16 ms, respectively. The number of frequencies $F$ is 256.
The resolution of candidate azimuths and elevations are both 5$^{\circ}$. %are 72 and 37, respectively.
%The maximum number of active sources $K_{\rm max}$ is set to 2.
During training, the length of microphone signals is set to 20 s, and correspondingly the number of time frames $N$ is 1249 which is pooled to be 104 by the network.
The model is trained using the Adam optimizer. %with mini-batches of 512 time frames
%with a learning rate of 0.0001.
The size of mini-batch is set to 66 (equals the number of microphone pairs).
%Each epoch contains 585 trajectories.
%The accuracy of azimuth and elevation estimation is separately evaluated.
The performance for the voice-active periods is evaluated with three metrics. The source is considered to be successfully localized if the azimuth error is not larger than 30$^{\circ}$. The mean absolute error (MAE) is computed by averaging the absolute azimuth (or elevation) error %(in spherical coordinates)
of all successfully localized sources and time frames. %for azimuth estimation and 15$^{\circ}$ for elevation estimation.
Miss detection (MD) refers to source active but not detected, and false alarm (FA) means source detected but not active.
The MD rate (MDR) and the FA rate (FAR) are computed as the percentage of MDs and FAs out of the active sources of all time frames, respectively. %, are also used for evaluation.

\vspace{-0.2cm}
\subsection{Experimental Results}
\vspace{-0.1cm}

\begin{figure}[t]
  \centering
  \includegraphics[width=0.96\linewidth]{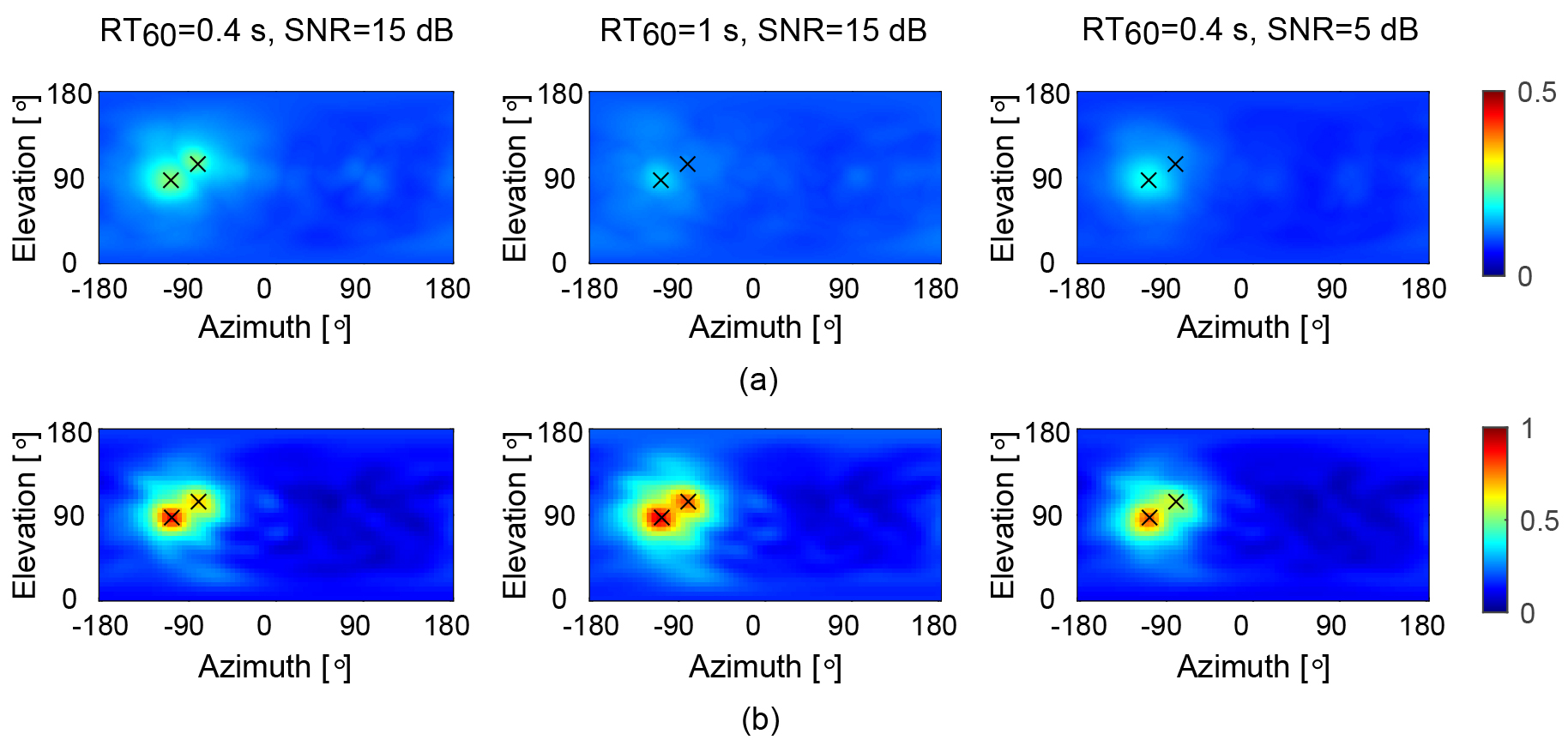}
  \vspace{-0.35cm}
  \caption{Illustration of spatial spectra of (a) SRP-PHAT \cite{LOCATA_SRP} and (b) SRP-DNN for two static sound sources present in the simulated rooms with different levels of noise and reverberation. The black crosses indicate the actual DOAs of sound sources.}
  \label{fig:exp_ss}
  \vspace{-0.4cm}
\end{figure}

%\begin{table}[t]
%  \caption{Performance for different settings of SPR-PHAT} %(Medium)
%  \label{tab:snrrt}
%  \vspace{-0.2cm}
%  \centering
%  \renewcommand\arraystretch{1.1}
%  \tabcolsep0.05in
%  %\small
%  \footnotesize
%  \scriptsize
%  \begin{tabular}{ccccccccccccccccc}
%    \hline
%    \hline
%    \multirow{1}{*}{Method} &\multirow{1}{*}{RT$_{60}$} & \multirow{1}{*}{SNR}
%      &MD rate [$\%$] &FA rate [$\%$] &MAE (az.$|$el.)  [$^{\circ}$] \\
%    \hline
%     \multirow{3}{*}{w/o DPTC}
%     &0.4 s &15 dB &1.0 &1.0 &2.8 $|$ 2.5 \\
%     &1 s &15 dB  &6.3 & 6.3 &4.1 $|$ 4.3 \\
%     &0.4 s &5 dB &7.2 & 7.2 &\ 5.4 $|$ 4.9
%%     &2 &1s &15dB \\
%%     &2 &0.4s &0dB
%     \vspace{0.1cm} \\
%     \multirow{3}{*}{\textbf{w/ DPTC}}
%     &0.4 s &15 dB &0.3 &0.3 &2.2 $|$ 1.9\\
%     &1 s &15 dB &0.4 &0.4 &2.7 $|$ 2.2\\
%     &0.4 s &5 dB &1.9 &1.9 &4.0 $|$ 3.6 \\
%%     &2 &1s &15dB \\
%%     &2 &0.4s &0dB \\
%    \hline
%    \hline
%  \end{tabular}
%\end{table}
%\subsubsection{Spatial spectrum estimation}
%\vspace{-0.1cm}
%To further investigate the assistance by the direct-path temporal context information,
%%we compare DOA estimation results of SRP-PHAT and DP-SPR-PHAT under different acoustic conditions.
%we estimate the spatial spectra using SRP-PHAT \cite{LOCATA_SRP} and Deep-SPR respectively, and plot them in the simulated two-source scenarios
Since the quality of spatial spectrum is important for the localization performance, we first visualize the spatial spectra of the proposed SRP-DNN method and one SRP-PHAT method \cite{LOCATA_SRP}. Fig.~\ref{fig:exp_ss} shows the spatial spectra obtained in the simulated rooms with different levels of noise and reverberation, where two static sources are present.
To compute the spatial spectrum, SRP-PHAT estimates the PHAT-weighted cross-power spectrum using Eq.~\eqref{eq_PHAT}, while SRP-DNN predicts the direct-path IPDs in Eq.~\eqref{eq_dp_PHAT} via DNN.
%As SRP-PHAT originally cannot deal with unknown source number cases, we replace the spectrum maximization in SPR-PHAT by peak detection \cite{Yan17}.
%which searches the significant spatial spectrum peaks that are larger than a predefined threshold.
%Fig.~\ref{fig:exp_ss} shows some examples of the spatial spectra estimated with baseline-SRP-PHAT and DPTC-SPR-PHAT respectively.
%The number of source is unknown.
Under the condition that RT$_{60}$=0.4 s and SNR=15 dB, both methods exhibit
sharp and distinct peaks around the actual DOAs. %which correspond to the direct-path propagated signals. %Deep-SPR exhibits relatively shaper and more distinct peaks around the actual DOAs than SRP-PHAT.
When the acoustic condition becomes worse, the local peaks of the proposed method are preserved, while the peaks of SRP-PHAT become flat and indistinctive. % when the acoustic condition become worse.
%and make the to exhibit sharper peaks at the actual DOAs even.
%This example is for single time interval in a simulated environment where sources are located at .
The robustness of SRP-DNN is mainly attributed to the fact that
the direct-path IPDs are well recovered by the CRNN and meanwhile the contamination of noise and reverberation is largely reduced. %which is mainly attributed to the fact that .

%we localize sources with baseline-SRP-PHAT and DPTC-SPR-PHAT respectively, and present the detailed results in the simulated single-source scenarios with different levels of noise and reverberation in Table~\ref{tab:snrrt}. The number of source is known. Comparing the performance of baseline-SRP-PHAT and DPTC-SPR-PHAT, it can be seen that exploiting the direct-path temporal context can significantly reduces the three metrics under all acoustic conditions. Tough both high-level noise and reverberation will degrade the localization performance, the degradation is relatively smaller for DPTC-SRP-PHAT. It confirms that direct-path temporal context is extremely helpful to improve the robustness of moving source localization under worse acoustic conditions.
%tends to make sources miss-detected, which is more obvious for SRP-PHAT. Comparing DPTC-SPR-PHAT with SRP-PHAT, it can be inferred that selecting a proper $\rho_{\rm TH}$ to guarantee a good trade-off between MD and FA under all the acoustic conditions is more difficult for SPR-PHAT. With the assistance of the direct-path temporal context, the proposed DPTC-SPR-PHAT method achieves more consistent and stable performance under different conditions.
%

%\vspace{-0.2cm}
%\subsubsection{DOA estimation}
%\vspace{-0.1cm}
Three baseline approaches are compared with the proposed SRP-DNN method on the LOCATA dataset, including Cross3D \cite{Cross3D21}, CTF-DPRTF \cite{LXF19} and SRP-PHAT \cite{LOCATA_SRP}.
Cross3D first computes SRP-PHAT and then tracks single sound source by performing 3D CNN on the sequence of the SRP-PHAT spatial spectrum.
CTF-DPRTF estimates the azimuths of multiple moving sources with the predicted DP-RTFs.
%SRP-PHAT localizes single sound source based on the correlations between microphone signals.
For the proposed SRP-DNN method, we present the results with either peak detection (PD) or iterative source detection and localization (IDL).
%Both Baseline-SRP-PHAT and DPTC-SRP-PHAT estimates spatial spectrum according to Eq.~\eqref{eq_SRP-PHAT}-\eqref{eq_GCC-PHAT}. But Baseline-SRP-PHAT predicts the PHAT cross-power spectrum by Eq.~\eqref{eq_PHAT} and averagely pools the temporal spatial spectra to compress the output time frames, while DPTC-SRP-PHAT predicts Eq.~\eqref{eq_dp_PHAT} and pools time frames both by causal CRNN. We tried two counting and localization methods, namely directly
%counting distinct peaks (DC) and iterative source detection and localization (IDL).
%. estimates spatial spectrum.
For the single-source case, the number of source is assumed to be known with $K$=1.
%$K_{\rm max}$=1 and ${\beta}_{\rm TH}$=-inf for the proposed method.
For the multi-source case, the number of source is assumed to be unknown. $K_{\rm max}$ and ${\beta}_{\rm TH}$ are separately set to 2 and 0.2 for SRP-DNN. The setting of ${\beta}_{\rm TH}$ is based on some preliminary experiments to seek a good trade-off between MDR and FAR.

\begin{table}[t]
 \caption{Performance of different methods on the LOCATA dataset}
  \label{tab:locata}
  \centering
  \renewcommand\arraystretch{1.09}
  \tabcolsep0.032in
  %\small
  %\footnotesize
  \scriptsize
  \begin{tabular}{cccccccccccccc}
    %\hline
    \hline
    \multirow{3}{*}{Method}  &\multicolumn{2}{c}{Single-source (task 3, 5)} & &\multicolumn{3}{c}{Multi-source (task 4, 6)} \\%&&\multicolumn{3}{c}{Avg.}\\
    \cline{2-3} \cline{5-7}
    %& &MD/FAR [$\%$] &MAE (az.$|$el.) [$^{\circ}$]  &  &MDR [$\%$] &FAR [$\%$] &MAE (az.$|$el.) [$^{\circ}$]  \\
    &MDR/FAR &MAE (az.$|$el.) &  &MDR &FAR  &MAE (az.$|$el.) \\
    &[$\%$] &[$^{\circ}$]  &  &[$\%$] &[$\%$] &[$^{\circ}$]  \\
    \hline
    %\multirow{1}{*}{TC \cite{TempContext21}} \\
    \multirow{1}{*}{Cross3D \cite{Cross3D21}}
    &0.9 &4.9 $|$ 3.3 &
    &-&-&-\ $|$\ - \\
    \multirow{1}{*}{CTF-DPRTF \cite{LXF19}}
    &2.4 &3.8 $|$ \ -\ \; &
    &17.6 &5.8 &4.8 $|$ \ - \; \\
    \multirow{1}{*}{SRP-PHAT} \cite{LOCATA_SRP}
    &0.8 &\textbf{2.5} $|$ \textbf{2.5} &
    &25.5 &12.1 &\textbf{2.3} $|$ 4.0  \\
    \hline
    \multirow{1}{*}{SRP-DNN+PD}
    &\textbf{0.1} &\textbf{2.5} $|$ 2.7 &
    &12.5 &7.9 &2.8 $|$ \textbf{3.7}  \\
    \multirow{1}{*}{\textbf{SRP-DNN+IDL (prop.)}}
    &\textbf{0.1} &\textbf{2.5} $|$ 2.7 &
    &\textbf{7.4}  &\textbf{4.0} &2.8 $|$ \textbf{3.7} \\
    %\hline
    \hline
   \end{tabular}
   % \scriptsize
%    \tabcolsep0.25in
%    \begin{tabular}{l}
%    $^*$: As Crooss3D cannot determine whether the source is active,  only the DOA estimates of source-active periods are used for performance measurement.
%    \end{tabular}
\vspace{-0.15cm}
\end{table}

\begin{figure}[t]
  \centering
  \includegraphics[width=1\linewidth]{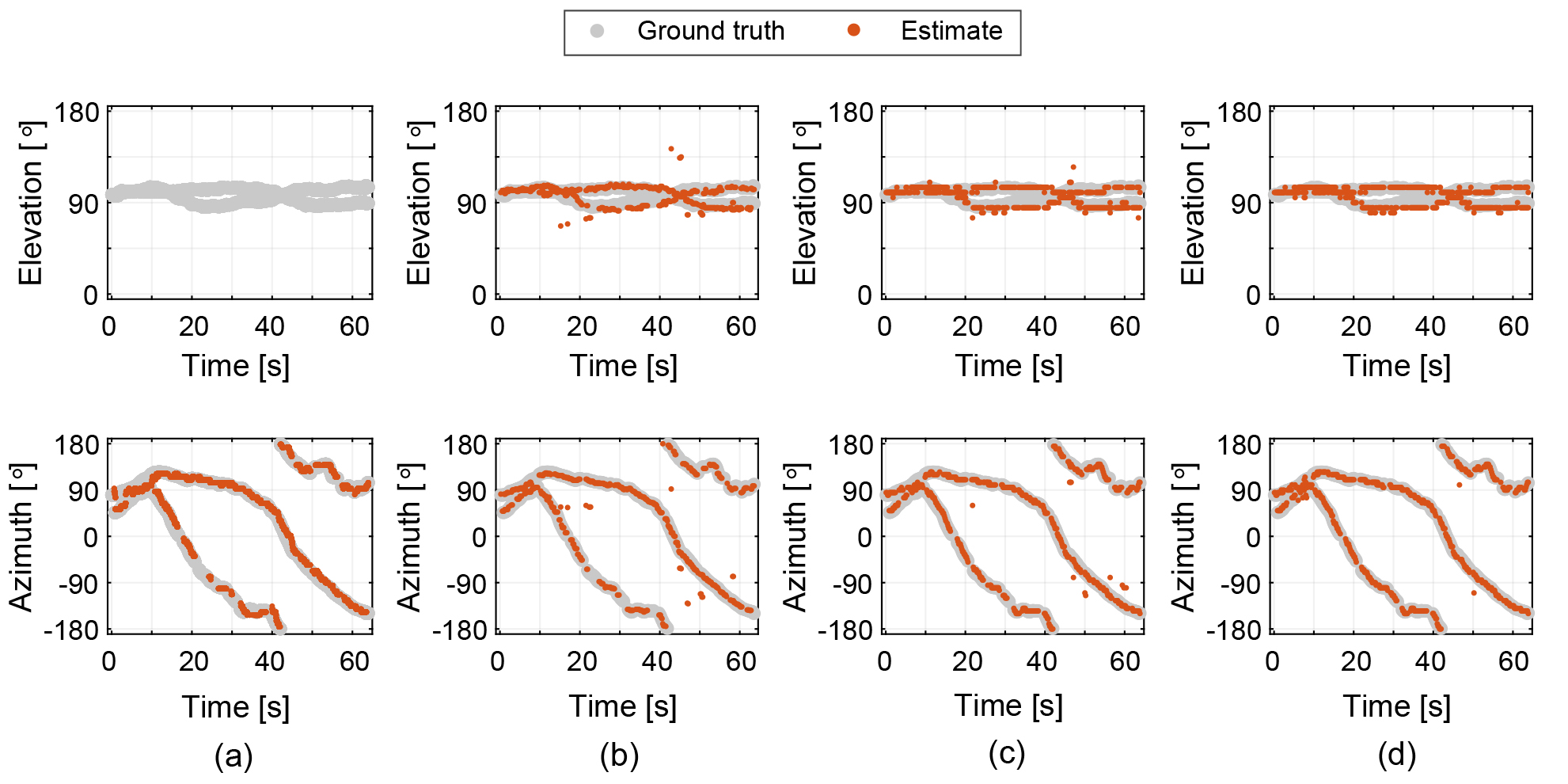}
  \vspace{-0.7cm}
  \caption{Illustration of DOA (trajectory) estimation of (a) CTF-DPRTF \cite{LXF19}, (b) SRP-PHAT \cite{LOCATA_SRP}, (c) SRP-DNN+PD and (d) SRP-DNN+IDL for one recording from the LOCATA challenge dataset. Two moving sound sources are present in this environment.} %Only the ground-truth   and the estimated active sources are shown.}
  \label{fig:exp_locata}
  \vspace{-0.4cm}
\end{figure}

The localization results are shown in Table~\ref{tab:locata}.
%It can be observed that
For the single-source case with known source number, the MDR and FAR are equal, and PD and IDL work in the same way.
SRP-DNN outperforms the other methods, and its MDR/FAR is extremely low. This indicates that SRP-DNN is able to largely reduce the influence of noise and reverberation, as which may cause spurious peaks.
For the multi-source case,
SRP-DNN achieves similar MAE and much smaller MDR and FAR compared to SRP-PHAT. This verifies that the proposed training target, i.e. Eq.~\eqref{eq_target}, can well model/preserve the direct-path IPD information of multiple sources, moreover the proposed CRNN model is able to well extract the target vector from microphone signals.
Relative to the trivial peak detection method, the proposed IDL method further improves the performance by disentangling the interaction of multiple sources.
SRP-DNN+IDL also performs better than CTF-DPRTF on all metrics.
%Under both settings of source number, the combination ofDPTC-SRP-PHAT and IDL achieves the best DOA estimation performance among different spatial spectrum estimation and source counting methods, and reduces approximately 16$\%$ MD rate and 13$\%$ FA rate than Baseline-SRP-PHAT+DC. This phenomenon demonstrates the effectiveness of direct-path temporal context assisted spatial spectrum and iterative source detection and localization algorithm.
Fig.~\ref{fig:exp_locata} presents an example of localizing two moving sources.
It can be seen that SRP-DNN+IDL provides relatively less erroneous DOA estimates (outliers) and more correct estimates, which is generally consistent with the lower rate of FA and MD presented in Table~\ref{tab:locata}.

\vspace{-0.2cm}
\section{Conclusion}
\vspace{-0.2cm}
This work proposes to learn competing and time-varying direct-path phase differences for robust multiple moving sound source localization. The designed causal CRNN fully exploits the TF patterns to learn the direct-path features which encodes both direct-path phase difference and source activity.
Using the predicted direct-path features, the SRP-DNN spatial spectrum shows more clear peaks around actual DOAs of sources even in the adverse noisy and reverberate scenarios.
By iteratively detecting and localizing the dominant source, the merged peaks of spatial spectrum can be separated, and accordingly the interaction between sources is reduced. Experimental results show the advantage of the proposed method over other methods for azimuth and elevation estimation of multiple moving sources in both simulated and real-world environments.
%
%The detailed performance comparison confirms the advantage of the direct-path feature extraction and the iterative source detection and localization for DOA estimation of moving sound sources.

\vfill\pagebreak

\bibliographystyle{IEEEbib}
\bibliography{mybib}
%\bibliography{strings,refs}

\end{document}